\documentclass[12pt]{article}
\usepackage{graphicx,amssymb,amsfonts,amsmath,epsfig,color}
\usepackage{mathrsfs}
\usepackage{cite}
\usepackage{url}
\newcounter{mnotecount}%[section]

\newcommand{\mnotex}[1]%{}
{\protect{\stepcounter{mnotecount}}$^{\mbox{\footnotesize $\bullet$\themnotecount}}$
\marginpar{%\color{red}%
\raggedright\tiny\em
$\!\!\!\!\!\!\,\bullet$\themnotecount: #1} }

\makeatletter
\DeclareSymbolFont{AMSb}{U}{msb}{m}{n}
\DeclareSymbolFontAlphabet{\mathbb}{AMSb}
\renewcommand{\section}{\@startsection{section}{1}{\z@}%
                                    {-7ex \@plus -1ex \@minus -.2ex}%
                                    {2.5ex \@plus.2ex}%
                                    {\normalfont\large\scshape\centering}}
\renewcommand{\subsection}{\@startsection{subsection}{2}{\z@}%
                                       {-5ex \@plus -1ex \@minus -.2ex}%
                                       {1.5ex \@plus.2ex}%
                                       {\normalfont\normalsize\scshape}}
\renewcommand{\subsubsection}{\@startsection{subsubsection}{3}{\z@}%
                                       {-5ex \@plus -1ex \@minus -.2ex}%
                                       {1.5ex \@plus.2ex}%
                                       {\normalfont\normalsize\scshape}}

\renewcommand\@seccntformat[1]{\ignorespaces\csname #1name\endcsname\space
                               \csname the#1\endcsname.\quad}   % Extra period and name added
\newdimen\captionmargin
\newdimen\captionindent
\newdimen\captionwidth
\newcommand{\captionfont}{\slshape}
\newcommand\@captionlabel[1]{\textsc{#1:}\space}
\long\def\@makecaption#1#2{%
  \vskip\abovecaptionskip
  \captionwidth\hsize
  \advance\captionwidth -2\captionmargin
  \sbox\@tempboxa{\@captionlabel{#1}\captionfont #2}%
  \ifdim \wd\@tempboxa >\captionwidth
    \ifdim\captionindent>\z@
      \advance\captionwidth -\captionindent
      \hskip\captionindent
    \fi
    \hskip\captionmargin
    \parbox[t]{\captionwidth}{\leavevmode\hskip-\captionindent
      \@captionlabel{#1}\captionfont #2}%
  \else
    \global \@minipagefalse
    \hb@xt@\hsize{\hfil\box\@tempboxa\hfil}%
  \fi
  \vskip\belowcaptionskip}
\def\eqnarray{%
   \stepcounter{equation}%
   \def\@currentlabel{\p@equation\theequation}%
   \global\@eqnswtrue
   \m@th
   \global\@eqcnt\z@
   \tabskip\@centering
   \let\\\@eqncr
   $$\everycr{}\halign to\displaywidth\bgroup
       \hskip\@centering$\displaystyle\tabskip\z@skip{##}$\@eqnsel
      &\global\@eqcnt\@ne$\;\hfil{##}$\hfil
      &\global\@eqcnt\tw@$\;\displaystyle{##}$\hfil\tabskip\@centering
      &\global\@eqcnt\thr@@ \hb@xt@\z@\bgroup\hss##\egroup
         \tabskip\z@skip
      \cr}
\ifcase \@ptsize
\or
\or
\fi
  \divide\@tempdima\baselineskip
  \@tempcnta=\@tempdima
\makeatother
\begin{document}

\renewcommand{\theequation}{\arabic{section}.\arabic{equation}}
\renewcommand{\thefigure}{\arabic{figure}}
\newcommand{\gapprox}{%
\mathrel{%
\setbox0=\hbox{$>$}\raise0.6ex\copy0\kern-\wd0\lower0.65ex\hbox{$\sim$}}}
\textwidth 165mm \textheight 220mm \topmargin 0pt \oddsidemargin 2mm
\def\ib{{\bar \imath}}
\def\jb{{\bar \jmath}}

\newcommand{\ft}[2]{{\textstyle\frac{#1}{#2}}}
\newcommand{\be}{\begin{equation}}
\newcommand{\ee}{\end{equation}}
\newcommand{\bea}{\begin{eqnarray}}
\newcommand{\eea}{\end{eqnarray}}
\newcommand{\Identity}{{1\!\rm l}}% Unit Matrix
\newcommand{\cx}{\overset{\circ}{x}_2}
\def\CN{$\mathcal{N}$}
\def\CH{$\mathcal{H}$}
\def\hg{\hat{g}}
\newcommand{\bref}[1]{(\ref{#1})}
\def\espai{\;\;\;\;\;\;}
\def\zespai{\;\;\;\;}
\def\avall{\vspace{0.5cm}}
\newtheorem{theorem}{Theorem}
\newtheorem{acknowledgement}{Acknowledgment}
\newtheorem{algorithm}{Algorithm}
\newtheorem{axiom}{Axiom}
\newtheorem{case}{Case}
\newtheorem{claim}{Claim}
\newtheorem{conclusion}{Conclusion}
\newtheorem{condition}{Condition}
\newtheorem{conjecture}{Conjecture}
\newtheorem{corollary}{Corollary}
\newtheorem{criterion}{Criterion}
\newtheorem{defi}{Definition}
\newtheorem{example}{Example}
\newtheorem{exercise}{Exercise}
\newtheorem{lemma}{Lemma}
\newtheorem{notation}{Notation}
\newtheorem{problem}{Problem}
\newtheorem{prop}{Proposition}
\newtheorem{rem}{{\it Remark}}
\newtheorem{solution}{Solution}
\newtheorem{summary}{Summary}
\numberwithin{equation}{section}
\newenvironment{pf}[1][Proof]{\noindent{\it {#1.}} }{\ \rule{0.5em}{0.5em}}
\newenvironment{ex}[1][Example]{\noindent{\it {#1.}}}

\thispagestyle{empty}

%\begin{flushright}\scshape
%\end{flushright}
%\vskip1cm

\begin{center}

{\LARGE\scshape Interiors of Singularity-Free Rotating Black Holes \par}
\vskip15mm

\textsc{Ram\'{o}n Torres\footnote{E-mail: ramon.torres-herrera@upc.edu}}
\par\bigskip
{\em
Dept. de F\'{i}sica, Universitat Polit\`{e}cnica de Catalunya, Barcelona, Spain.}\\[.1cm]
%

%\vspace{5mm}

\end{center}

\begin{abstract}
General Relativity provides us with some solutions for rotating black holes. However, there are some problems associated with them: the appearance of singularities, the possibility of violations of the cosmic censorship conjecture, the existence of regions where the mass acts repulsively and the violation of causality. Many authors consider that these problems reveal the existence of certain limits in the applicability of General Relativity. For instance, it is believed that the same existence of singularities in the classical black hole solutions is a weakness of the theory and that a full Quantum Gravity Theory would provide us with singularity-free black hole models. In this paper, the generic properties of the interiors of singularity-free rotating black holes are analyzed. Remarkably, it is shown that they are devoid of any of the aforementioned problems of the classical solutions.
\end{abstract}

\vskip10mm
\noindent KEYWORDS: Rotating black holes; Regular black holes; Quantum black holes.

\setcounter{equation}{0}

\section{Introduction}\label{secIntro}

Most astrophysically significant
bodies are rotating. If a rotating body collapses, the rate of rotation
will speed up, maintaining constant angular momentum. Through a rather complicated process, the body could finally generate a black hole which would be a \emph{rotating black hole} (\emph{RBH}).

Recently, there have been advances in the modelling of both non-singular (also known as \textit{regular}) rotating black holes coming from many different theoretical approaches.
%v2
(See, for example, some spherically symmetric models in \cite{A-BI}\cite{A-BII}\cite{B&V}\cite{Bardeen})\cite{A&B2005}\cite{B&R}
\cite{Frolov2014}\cite{G&P2014}\cite{Hay2006}\cite{H&R2014}\cite{dust2014} and references therein).
%v2end
Undoubtedly, the recent observational developments (LIGO-VIRGO-KAGRA collaboration, the Event Horizon Telescope or, in the near future, the LISA project) and the possibility to probe our theoretical predictions have greatly contributed to awakening interest.

From a classical point of view,
the spacetime corresponding to an uncharged rotating black hole is described by a Kerr solution.
Let us now briefly summarize the characteristics of its interior.
(The reader can consult, for example, \cite{Griff}\cite{gravit} and references therein for more information). In Boyer-Lidquist (B-L) coordinates $\{t,r,\theta,\phi\}$, the Kerr metric takes the form
\begin{equation}\label{gKerr}
ds^2=-\frac{\Delta}{\Sigma} (dt-a \sin^2\theta d\phi)^2+
\frac{\Sigma}{\Delta} dr^2+\Sigma d\theta^2+\frac{\sin^2\theta}{\Sigma}(a dt-(r^2+a^2)d\phi)^2,
\end{equation}
where
\[
\Sigma=r^2+a^2 \cos^2\theta, \hspace{1cm} \Delta=r^2-2 m r+a^2,
\]
$m$ is the black hole mass
and $a$ is a \textit{rotation parameter} that measures the (Komar) angular momentum per unit of mass \cite{gravit}. The spacetime is type D if $m\neq 0$.

If $m\neq 0$ there is a curvature singularity at $(r=0, \ \theta=\pi/2)$, as can be shown by the divergence of the curvature invariant $R_{\alpha\beta\gamma\delta} R^{\alpha\beta\gamma\delta}$.
Remarkably, for $a\neq 0$ and $\theta\neq \pi/2$, a surface defined by $t=$constant and $r=0$, known as \textit{the disk}, is singularity-free and has metric
\[
ds_{Disk}^2=a^2 \cos^2 \theta d\theta^2+a^2 \sin^2 \theta d\phi^2=dx^2+dy^2,
\]
where the coordinate change $x\equiv a \sin\theta \cos \phi$, $y\equiv a \sin\theta\sin\phi$ has been made to make explicit that the surface is flat.
The disk corresponds to $x^2+y^2<a^2$, while the curvature singularity corresponds to \textit{the ring} $x^2+y^2=a^2$.
In this way, the curves that reach $r=0$ with $\theta\neq \pi/2$ are reaching a regular point. In order to continue the curves through the disk it is usually argued that an analytic extension of the spacetime has to be performed.
The procedure requires letting the coordinate $r$ to take negative values \cite{H&E}. The $r<0$ extended spacetime can be seen as an asymptotically flat spacetime with negative mass. Causality violations occur in the extended spacetime \cite{Carter1968a}.

Several authors have suggested that the existence of singularities in the solutions of General Relativity has to be considered as a weakness of the theory rather than as a real physical prediction.
The problem of obtaining regular models for black holes was first approached for spherically symmetric black holes.
However, recently there have appeared different proposals
for \emph{regular} rotating black hole spacetimes with their corresponding metrics.
Some heuristic proposals can be found in \cite{A-A}\cite{B&M}\cite{FLMV}\cite{GM}\cite{LGS}\cite{Maeda}\cite{MFL}\cite{eye}), where their regular interiors are analyzed with different procedures.
Some authors, inspired by the work of Bardeen \cite{Bardeen}, have taken the path of nonlinear electrodynamics, which provides the necessary modifications in the energy-momentum tensor in order to avoid singularities in the RBH \cite{D&G}\cite{GD}\cite{Ghosh}\cite{Lamm}\cite{Tosh}\cite{TSA}. Yet, another way of addressing the problem of singularities is to take into account that Quantum Gravity effects should play an important role in the core of black holes, so that it would seem convenient to directly derive the black hole behaviour from an approach to Quantum Gravity. In this way, regular RBHs deduced in the Quantum Einstein Gravity approach can be found in \cite{R&T}\cite{TorresExt}, in the framework of Conformal Gravity in \cite{BMR}, in the framework of Shape Dynamics in \cite{G&H}, inspired by Supergravity in \cite{Buri}, by Loop Quantum Gravity in \cite{C&M} and by non-commutative gravity in \cite{S&S}.

In this article we show that regular rotating black holes never need extensions through their \textit{disk}. In this way, there is no need for regions with repulsive masses. Furthermore,
causality violations can be naturally avoided.

The article is divided as follows.
In section \ref{KRBHs} the metric for singularity-free rotating black holes is introduced and their main features are analyzed. Note that, even if a regular RBH model could usually come from an approach to Quantum Gravity Theory, here it is assumed that it can be reasonably well described by a manifold endowed with its corresponding metric.
In section \ref{secr0}, the need for extensions through $r=0$ in Kerr's solution is contrasted with the situation for regular RBH. Section \ref{ring} is devoted to the analysis of the (regular) ring, the absence of conical singularities and its directional character. In section \ref{secCaus}, it is shown how causality problems can be avoided in regular RBHs. The global structure of these spacetimes, which will necessarily differ from the maximally extended Kerr's solution, is treated in section \ref{Horizons}. Finally, section \ref{conclu} is devoted to the conclusions.

\section{Kerr-like Rotating Black Holes}\label{KRBHs}

While rotating black holes have been obtained by different approaches, most of them share a common \textit{Kerr-like form}.
The general metric corresponding to this kind of RBH, was found by G\"{u}rses-G\"{u}rsey \cite{GG} as a particular rotating case of the algebraically special Kerr-Schild metric:
\begin{equation}\label{GGg}
ds^2=(\eta_{\alpha \beta} +2 H k_\alpha k_\beta) dx^\alpha dx^\beta,
\end{equation}
where $\eta$ is Minkowski's metric, $H$ is a scalar function and $\mathbf k$ is a light-like vector both with respect to the spacetime metric and to Minkowski's metric.
Specifically, in Kerr-Schild (K-S) coordinates $\{\tilde{t},x,y,z\}$ the  G\"{u}rses-G\"{u}rsey metric (\ref{GGg}) corresponds with the choices
\begin{equation}\label{HDef}
H=\frac{\mathcal M (r) r^3}{r^4+a^2 z^2}
\end{equation}
and
\begin{equation}\label{kDef}
k_\alpha dx^\alpha =-\frac{r (x dx+y dy) -a (x dy-y dx)}{r^2+a^2}-\frac{z dz}{r}-d\tilde{t},
\end{equation}
where $r$ is a function of the Kerr-Schild coordinates implicitly defined by
\begin{equation}\label{defr}
r^4-r^2 (x^2+y^2+z^2-a^2) -a^2 z^2 =0,
\end{equation}
$\mathcal M (r) $ is known as the \textit{mass function}
and the constant $a$ is a rotation parameter.

This metric can be written in Boyer-Lindquist-like coordinates by using the coordinate change defined by
\begin{eqnarray*}
x+i y&=&(r+i a) \sin\theta\exp\left[i\int(d\phi+\frac{a}{\Delta}dr)\right]\\
z&=&r \cos\theta\\
\tilde{t}&=&t+\int \frac{r^2+a^2}{\Delta} dr-r.
\end{eqnarray*}
where now $\Delta=r^2-2 \mathcal M(r) r+a^2$.
The resulting metric takes the form
\begin{equation}\label{gIKerr}
ds^2=-\frac{\Delta}{\Sigma} (dt-a \sin^2\theta d\phi)^2+
\frac{\Sigma}{\Delta} dr^2+\Sigma d\theta^2+\frac{\sin^2\theta}{\Sigma}(a dt-(r^2+a^2)d\phi)^2,
\end{equation}
where, again, $\Sigma=r^2+a^2 \cos^2\theta$.
Note that this metric reduces to Kerr's solution in B-L coordinates if $\mathcal M(r)=m$=constant.

As in Kerr's solution, for $\theta\neq\pi/2$, the surface $t=$ constant and $r=0$ is a flat surface (corresponding to $x^2+y^2<a^2$) that will be called \textit{the disk}. Likewise, \textit{the ring} corresponds to $\theta=\pi/2$ (or $x^2+y^2=a^2$).

Note also the symmetry $\{a,\phi\}\leftrightarrow \{-a,-\phi\}$ in this metric. This allows us, for the sake of simplicity, to assume $a>0$ in this article, since the negative case is covered by using the trivial coordinate change $\phi \rightarrow -\phi$.

In order for the model of a RBH to be regular it should be devoid of curvature singularities.
We say that there is a \textit{scalar curvature singularity}
in the spacetime if any scalar invariant polynomial in the Riemann tensor diverges when approaching it along any incomplete curve.

Scalar curvature singularities may appear if $\Sigma=0$ or, in other words, in $(r=0,\theta=\pi/2)$. (We already confirmed this possibility in the introduction for the particular case of Kerr's solution).
Now, by explicitly computing the complete set of scalars in our case, one directly gets a necessary and sufficient condition for the absence of scalar curvature singularities:

\begin{theorem}\label{teorema}\cite{TorresReg}
Assuming a RBH metric (\ref{gIKerr}) possessing a $C^3$ function $\mathcal M(r)$, all its second order curvature invariants will be finite at $(r=0,\theta=\pi/2)$ if, and only if,
\begin{equation}\label{condisreg}
 \mathcal M (0)= \mathcal M' (0)= \mathcal M'' (0)=0 .
\end{equation}
\end{theorem}

%v2
Many proposals of regular RBH with mass functions satisfying the conditions in the theorem have appear in the literature. Note that any function admitting an expansion around $r=0$ in the form $\mathcal M(r)=c r^n + O(r^{n+1})$, with $c$ a constant and $n\geq 3$, will satisfy the conditions. (This will be written here also as $\mathcal M(r)= O( r^n )$ or $\mathcal M(r)\sim r^n$). Consider, for example, the rotating versions \cite{B&M} of the Bardeen \cite{Bardeen} and Hayward \cite{Hay2006} regular spherically symmetric black holes with mass functions
\begin{equation}\label{B&H}
\mathcal M(r)_{Bardeen}=\frac{m r^3}{(r^2+l^2)^{3/2}} \ \ ; \ \ \mathcal M(r)_{Hayward}=\frac{m r^3}{r^3+l^3},
\end{equation}
where $l$ is a constant. These mass functions admit the following expansions around $r=0$:
\[
\mathcal M(r)_{Bardeen}=\frac{m}{l^3} r^3 + O(r^5) \ \ ; \ \ \mathcal M(r)_{Hayward}=\frac{m}{l^3} r^3 + O(r^6).
\]
%v2end

In order to analyze the general properties of the RBH spacetime
we will use the following null tetrad-frame:
\begin{eqnarray*}
\mathbf{l} &=&\frac{1}{\Delta} \left( (r^2+a^2) \frac{\partial}{\partial t}+\Delta \frac{\partial}{\partial r}+a \frac{\partial}{\partial \phi}\right),\\
\mathbf n &=&\frac{1}{2 \Sigma} \left( (r^2+a^2) \frac{\partial}{\partial t}-\Delta \frac{\partial}{\partial r}+a \frac{\partial}{\partial \phi}\right),\\
\mathbf m &=& \frac{1}{ \sqrt{2} \varrho } \left(i a \sin\theta \frac{\partial}{\partial t}+\frac{\partial}{\partial \theta}+i \csc\theta \frac{\partial}{\partial \phi} \right),\\
\mathbf{\bar m} &=& \frac{1}{ \sqrt{2} \bar\varrho } \left(-i a \sin\theta \frac{\partial}{\partial t}+\frac{\partial}{\partial \theta}-i \csc\theta \frac{\partial}{\partial \phi} \right),
\end{eqnarray*}
where $\varrho\equiv r+i a \cos\theta$, $\bar\varrho\equiv r-i a \cos\theta$ and the tetrad is normalized as follows $\mathbf l^2=\mathbf n^2=\mathbf m^2=\mathbf{\bar m}^2=0$ and $\mathbf l\cdot \mathbf n=-1= -\mathbf{m}\cdot \mathbf{\bar m}$.

It can be shown \cite{TorresReg} that the RBH metric (\ref{gIKerr}) with $\mathcal M(r)\neq 0$ is Petrov type D and that the two double principal null directions are $\mathbf l$ and $\mathbf n$.

We can also define a real orthonormal basis $\{\mathbf{t}, \mathbf{x}, \mathbf{y}, \mathbf{z } \}$
formed by a timelike vector $\mathbf{t}\equiv (\mathbf{l}+\mathbf{n})/\sqrt{2}$ and three spacelike vectors: $\mathbf{z}\equiv (\mathbf{l}-\mathbf{n})/\sqrt{2}$,  $\mathbf x=(\mathbf m +\bar{\mathbf m})/\sqrt{2}$ and $\mathbf y=(\mathbf m -\bar{\mathbf m}) i/\sqrt{2}$. Then,
$\mathbf t$ and $\mathbf z$ are two eigenvectors of the Ricci tensor with eigenvalue \cite{TorresReg}
\begin{equation}\label{lambda1}
\lambda_1=\frac{2 a^2 \cos^2{\theta} \mathcal M'+r \Sigma \mathcal M''}{\Sigma^2}.
\end{equation}
$\mathbf x$ and $\mathbf y$ are two eigenvectors of the Ricci tensor with eigenvalue
\begin{equation}\label{lambda2}
\lambda_2=\frac{2 r^2 \mathcal M'}{\Sigma^2}.
\end{equation}

In this way, the Ricci tensor can be written as
\begin{equation}\label{Ricci}
R_{\mu\nu}= \lambda_1\, (-t_\mu t_\nu+z_\mu z_\nu)+ \lambda_2 (x_\mu x_\nu+y_\mu y_\nu),
\end{equation}

Even if we are not confined to General Relativity we can consider the existence of an \textit{effective energy-momentum tensor} defined through
\[
T_{\mu\nu}\equiv R_{\mu\nu}-\frac{1}{2} \mathcal R g_{\mu\nu}.
\]

If we take the expression obtained for the Ricci tensor (\ref{Ricci}) one can explicit $\mathbf{T}$ for a RBH as
\[
T_{\mu\nu}=-\lambda_2 (-t_\mu t_\nu + z_\mu z_\nu)- \lambda_1 (x_\mu x_\nu+ y_\mu y_\nu).
\]
Since $\mathbf{T}$ diagonalizes in the orthonormal basis $\{\mathbf{t}, \mathbf{x}, \mathbf{y}, \mathbf{z } \}$, the RBH spacetime possesses an (effective) energy-momentum tensor of type I \cite{H&E}. The (effective) density being $\mu=\lambda_2$ and the (effective) pressures being $p_x=p_y=-\lambda_1$ and $p_z=-\lambda_2$.

%v2
The weak energy conditions \cite{H&E} require $\mu\geq 0$ and $\mu+p_i\geq 0$. In other words, in this case they require
\[
\lambda_2 \geq 0 \hspace{1 cm} \mbox{and} \hspace{1 cm} \lambda_2-\lambda_1\geq 0.
\]
By using this and expressions (\ref{lambda1}) and (\ref{lambda2}) it is easy to show the following
\begin{prop}\cite{TorresReg}
Assume that a \emph{regular} RBH has a function $\mathcal M (r)$ that can be approximated by a Taylor polynomial around $r=0$, then the weak energy conditions should be violated around $r=0$.
\end{prop}

Note that for this type I effective energy-momentum the violation of the weak energy condition also implies the violation of the \textit{dominant} and the \textit{strong} energy conditions.

In this way, no model with \textit{normal} matter (matter satisfying the energy conditions) can produce a \emph{regular} rotating black hole of the type (\ref{gIKerr}).
However, the violation of the WEC around $r=0$ is not problematic since it is well-known that quantum effects can violate the WEC (Casimir effect). Moreover, singularity theorems require the spacetime to fulfill some energy condition in order to predict the existence of singularities. In this sense, the violation of energy conditions just \textit{helps} to avoid the existence of singularities.
%v2end

\section{Extensions through the disk}\label{secr0}

As stated in the introduction, for Kerr's solution
one should extend the spacetime through the disk.
Now, in order to analyze the general situation for regular RBHs with metric (\ref{GGg}), let us proceed with an analysis similar to the one usually carried out for the classical RBH case. Consider the metric component
\begin{equation}\label{gtt}
g_{tt}=-1+ \frac{2 \mathcal M (r) r^3}{r^4+a^2 z^2}.
\end{equation}
Let us imagine and observer crossing $r=0$ moving in the $z$ \textit{axis} ($x=y=0$).
If we choose $r$ to be non-negative, then (\ref{defr}) implies that $r=|z|$ along the trajectory of the observer, so that along it
\begin{equation}\label{gttz}
g_{tt}=-1+ \frac{2 \mathcal M (|z|) |z|}{z^2+a^2}.
\end{equation}
The numerator in the fraction indicates that the derivative of this metric component along the axis, as well as the Christoffel symbols and the extrinsic curvature of the surface can be discontinuous across the disk depending on the chosen mass function. As was already mentioned, a well-known relevant case of this discontinuity occurs if the mass function is constant: Kerr's solution.

\begin{figure}[ht]
\includegraphics[scale=1.1]{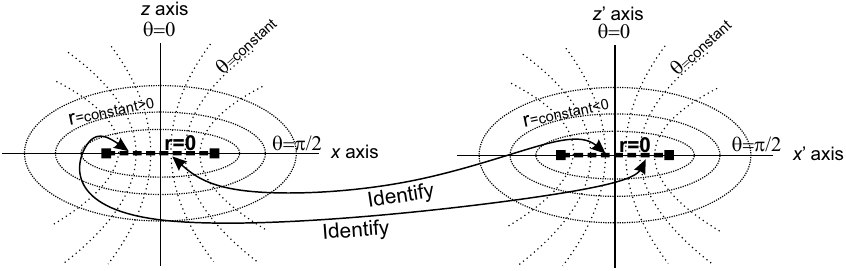}
\caption{\label{extensionr0} In Kerr's solution, the extension through $r=0$ is obtained by identifying the top of the surface ($r=0, t=$constant) in the hypersurface described by coordinates $\{x,y,z\}$ with the bottom of the surface ($r=0, t=$constant) in the hyersurface described by coordinates $\{x',y',z'\}$, and vice versa. Only the $y=0$, $y'=0$ sections of these hypersurfaces are represented here.}
\end{figure}

The differentiability problems in Kerr's RBH can be solved by analytically extending the spacetime through $r=0$ with negative values for $r$. This requires considering two spacetimes, one with positive $r$ and another with negative $r$ and properly identifying points in their $r=0$ surfaces by a standard procedure which is ilustrated in figure \ref{extensionr0} (see, for example, \cite{H&E})

At first sight, the situation for general regular RBH looks much better. Assuming that the regular RBH has a mass function $\mathcal M (r) \sim r^n$ with $n\geq 3$ around $r=0$, the metric component along the trajectory (\ref{gttz}) will not have differentiability problems in $z=0$ ($\partial_z g_{tt}(z=0)=0$) (and, in fact, it will be at least\footnote{Specifically, it will be $C^n$ if $n$ is even and $C^\infty$ if $n$ is odd.} $C^n$). This suggests that the extension through the disk could not be necessary for regular RBH.
In order to prove this conjecture, one has to go beyond a particular trajectory intersecting the disk and beyond the analysis of a single metric component. Let us start by noticing that while approaching a point in the disk ($x^2+y^2<a^2$), according to (\ref{defr}), the function $r$ approaches zero whenever $z$ approaches zero and vice versa. If one chooses to avoid an extension with $r<0$, we get, solving for $r$ in (\ref{defr}), that around $z=0$\footnote{Note that we assumed $a>0$ (at the beginning of this section) throughout the article. (If not, here we should replace $a\rightarrow |a|$).}
\[
r \simeq \frac{a}{\sqrt{a^2-(x^2+y^2)}} |z|
\]
If we introduce this into the metric component (\ref{gtt}) and consider a mass function $\mathcal M (r) \sim r^n$ with $n\geq 3$ around $r=0$, we see that the metric component takes the form
\[
g_{tt} \simeq -1+\frac{f(x,y) |z|^{n+1}}{g(x,y) z^2+a^2},
\]
where $f$ and $g$ are finite differentiable functions in the disk. In this way, $g_{tt}$ is differentiable at the disk. (Specifically, again ($\partial_z g_{tt}(z=0)=0$) and, in fact, $g_{tt}$ is at least $C^n$ at the disk).
The reader can easily check that a similar situation is found for the rest of metric components. Let us only remark that the metric will not be analytic at the disk independent of $n$. This is because not all metric components will be infinitely differentiable. For example, even if the particular metric component (\ref{gtt}) for odd $n$ is $C^\infty$ in the disk, other metric components like
\[
g_{xz}=\frac{2\mathcal{ M}(r) r^2 z (a y+x r)}{(a^2+r^2)(a^2 z^2+r^4)}
\]
are not. ($g_{xz}$ is $C^n$ for odd $n$). Nevertheless, since usually the metric is required to be at least $C^2$ \cite{H&E}\footnote{However, many authors consider even this degree of differentiability too high. See, for instance, \cite{Seno}\cite{FMS} and references therein.}, a $C^n$ metric with $n\geq 3$ at the disk is more than enough.

In summary, we have arrived at the following result:

\begin{em}
Regular black holes (without an extension through $r=0$) have a high degree of differentiability at the disk (at least $C^n$, with $n \geq 3$). In this way, regular RBHs do not have differentiability problems at the disk and an extension through $r=0$ with $r<0$ is not needed \footnote{Let us comment that, even if not mathematically needed, the possibility of extending through $r=0$ with negative values of $r$ exists, in principle, for all regular RBHs. Nevertheless, one finds in addition to the problems already commented with this approach in the classical solutions, new mathematical and physical problems \cite{TorresExt}.}.
\end{em}

\begin{figure}[ht]
\includegraphics[scale=0.7]{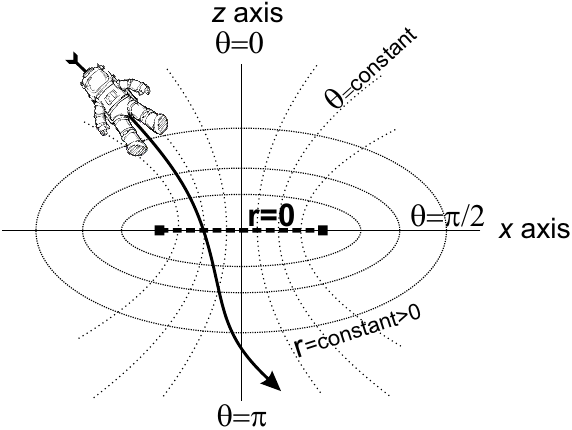}
\caption{\label{caseB} For regular RBHs no extension through $r=0$ is required. An observer crossing the surface ($r=0,t=$constant) from positive $z$ to negative $z$ following a non-geodesic time-like curve can stay in its original spacetime. Along the trajectory of the observer $r$ just decreases until reaching the surface $r=0$, where it increases again. The angular B-L coordinates just jump when crossing the disk. In the plane shown in the figure: $\theta\rightarrow \pi-\theta$. (This is somehow similar to following a trajectory crossing $r=0$ in a regular spherically symmetric spacetime equipped with spherical coordinates).}
\end{figure}

In this way,  $r$ would remain non-negative along the trajectory of an observer crossing through the disk, as explicitly shown in figure \ref{caseB}.

\section{The ring}\label{ring}

The ring itself ($z=0$ and $x^2+y^2=a^2$) requires a separate analysis. If it is approached with $x^2+y^2=a^2$ and we avoid extensions with $r<0$ then (\ref{defr}) implies that $r \simeq \sqrt{a |z|}$ around the ring. Consider, for example, a mass function $\mathcal M(r) = m_n r^n + O(r^{n+1})$. The metric coefficient (\ref{gtt}) around the ring behaves as
\[
g_{tt}\simeq -1+m_n (a |z|)^\frac{n-1}{2}.
\]
For a regular RBH theorem  \ref{teorema} requires $n\geq 3$. In this case this metric coefficient is continuous and differentiable at the ring for $n>3$, but it is only Lipschitz continuous ($C^{1-}$) for $n=3$.
(The reader can check that other metric coefficients in K-S coordinates have similar, or better, behaviours).

%In this way, for $n>3$ the behaviour of the ring does not seem to present any difficulties, while for $n=3$ the %behaviour of the ring loosely reminds that of the tip of a cone (what is a rather poor analogy as we will show %below).

%
%since neither the Kerr-Schild coordinates nor the Boyer-Lindquist coordinates are well-behaved in the ring. Take, for %instance, the metric in Kerr-Schild coordinates (\ref{GGg}) which depends on $H$ (\ref{HDef}) which ,(\ref{kDef}) and %(\ref{gIKerr})).
%

\begin{figure}[ht]
\includegraphics[scale=.8]{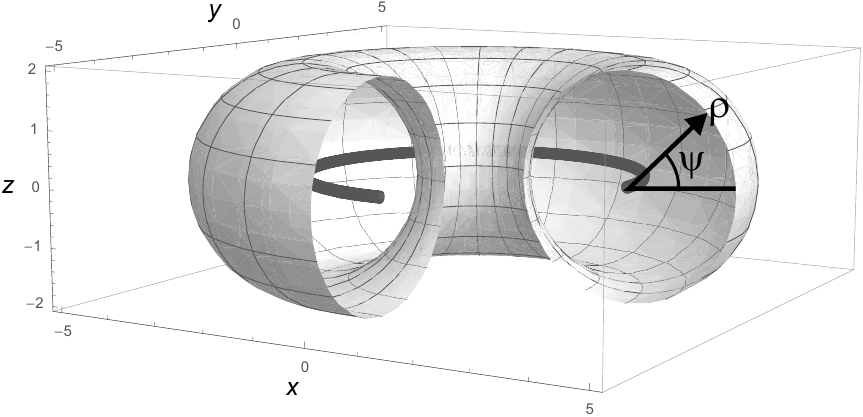}
\caption{\label{torusring} A plot of the behaviour of the toroidal coordinates with the (regular) ring ($\rho=0$) highlighted in black.}
\end{figure}

To study the ring in greater depth we will now introduce a set of coordinates that are better adapted to its structure and that will help us to analyze the directional behaviour of physical magnitudes around the ring: the \textit{toroidal} coordinates $\{\rho,\varphi,\psi\}$ (see figure \ref{torusring}). These coordinates are related to the Kerr-Schild coordinates through
\begin{eqnarray}\label{transf}
  x &=& (a+\rho \cos \psi) \cos \varphi,\nonumber \\
  y &=& (a+\rho \cos \psi) \sin \varphi, \\
  z &=& \rho \sin \psi. \nonumber
\end{eqnarray}
Note that in these coordinates the ring is defined by $\rho=0$.
%The Jacobian determinant of this transformation ($J=-\rho (a+\rho \cos \psi)$) implies that in these coordinates %there will be removable coordinate singularities both at the ring and at the $z$ axis.

%v2
%v2end

From the relationship between K-S and B-L coordinates, it follows that the coordinates $\{\rho,\psi\}$ are related to the Boyer-Lindquist coordinates $\{r,\theta\}$ through
\begin{eqnarray}\label{rcosth}
  r \cos \theta &=& \rho \sin \psi, \\
  (r^2+a^2) \sin^2\theta &=& (a+\rho \cos \psi)^2.\nonumber
\end{eqnarray}

If we do not perform an extension beyond $r=0$ with negative $r$, this implies that around the ring

\begin{eqnarray}\label{rthaprox}
  r &\simeq& \sqrt{a (1+\cos \psi) \rho},\\
  \cos \theta &\simeq& \frac{\rho^{1/2} \sin \psi}{\sqrt{a (1+\cos\psi)}}.\nonumber
\end{eqnarray}

Note that the metric (\ref{GGg})
can be interpreted as consisting of a Minkowskian part plus a \textit{perturbation}.
With regard to the perturbation, the behaviour of $H$ around the ring in toroidal coordinates is, from (\ref{rthaprox}),
\[
H=\frac{\mathcal M (r) r^3}{r^4+a^2 z^2}\simeq
\frac{\mathcal{M} (\rho, \psi)}{2} \sqrt{\frac{1+\cos\psi}{a \rho}}.
\]

In this way, for a regular RBH with $\mathcal M (r) \sim r^n$ ($n\geq 3$) around $r=0$ one has $H\propto \rho^{(n-1)/2}$ which tends to zero as the ring is approached.

On the other hand, around the ring (\ref{kDef}) can be written in toroidal coordinates as

\[
k_\alpha dx^\alpha \simeq-dt+a d\varphi- \sqrt{(1+\cos\psi)\frac{\rho}{a}} d\rho+\frac{\rho^{3/2}}{\sqrt{a (1+\cos \psi)}} \sin \psi d\psi.
\]

As a consequence, the leading order behaviour of the metric around the ring is just the leading order of the Minkowskian spacetime in toroidal coordinates:

\[
ds^2\simeq -dt^2+d\rho^2+\rho^2 d\psi^2+ a^2 d\varphi^2.
\]

Note that this metric is degenerated at $\rho=0$ since its determinant is zero there. This is expected from the behaviour of the Jacobian determinant of the transformation (\ref{transf}) ($J=-\rho (a+\rho \cos \psi)$).
%As expected, by not requiring the extension at the disk, the metric around the ring only has a removable coordinate %singularity at $\rho=0$ \footnote{

It is interesting to note that a curve around the ring with $\{ t$, $\rho=\delta\rho$,  $\varphi\}$ constants describes a circle of length $2\pi \delta \rho$. Therefore, the ring of a regular RBH is not a \textit{conical singularity}, as it is for the singular ring in Kerr spacetime \cite{G&V}\cite{Chrus}.
%However, if one insists in extending the metric for negative values of $r$, the curve should travel both the original %spacetime and the $r<0$ spacetime. Consequently, $psi$  coordinate $\psi$ would have required a range from 0 to $4 %\pi$, what would imply a \textit{conical singularity} at the ring.

Let us now analyze the behaviour of quantities that are not dependent on the coordinates: The second order curvature scalars (computed, by definition, using second order derivatives of the metric). We know that they are all finite if the conditions in theorem \ref{teorema} are fulfilled. Let us now analyze their continuity. Consider, for example, the curvature scalar in B-L coordinates:
\[
\mathcal R=\frac{2 (2 \mathcal M'(r)+r \mathcal M''(r))}{r^2+a^2 \cos ^2 \theta}.
\]
If the mass function behaves around the ring as $\mathcal M(r) = m_n r^n + O(r^{n+1})$, then the curvature
behaves around the ring as
\[
\mathcal R\simeq \frac{2 m_n n (1+n) r^{n-1}}{r^2+a^2 \cos ^2 \theta},
\]
which in toroidal coordinates can be written as
\[
\mathcal R\simeq m_n n (1+n) (1+\cos \psi)^\frac{n-1}{2} (a \rho)^\frac{n-3}{2}.
\]
In this way, for $n>3$ the curvature scalar is continuous (and zero) at the ring, however for $n=3$ it is finite, but not continuous at the ring, where it has a clear directional character ($\mathcal R\simeq 24 m_3 (1+\cos \psi)$).
It can be checked that the same behaviour is repeated for all second order scalar curvature invariants. (An algebraically independent set of them is listed in \cite{TorresReg}).

As stated in section \ref{KRBHs}, the effective density ($\mu$) and pressures ($p_x,\ p_y\ \mbox{and}\ p_z$) measured by observers with 4-velocity $\mathbf t$ can be directly obtained from $\lambda_1$ and $\lambda_2$.
Let us now analyze their dependence on $\rho$ around the ring for a mass function $\mathcal M(r)\sim r^n,\ (n\geq 3)$ around $r=0$. With the help of (\ref{rthaprox}):

\begin{eqnarray*}
\mu=-p_z&=&\lambda_2=\frac{2 r^2 \mathcal M'}{\Sigma^2} \sim \rho^\frac{n-3}{2}\\
p_x=p_y&=&-\lambda_1=-\frac{2 a^2 \cos^2{\theta} \mathcal M'+r \Sigma \mathcal M''}{\Sigma^2} \sim \rho^\frac{n-3}{2}.
\end{eqnarray*}

In this way, the effective density and pressures vanish at the ring for $n > 3$. On the other hand, for $n=3$ the densities and pressures are finite functions of the direction of approach to the ring $\psi$. Only in this case the density and pressures are not continuous at the ring. Specifically, for $\mathcal M(r)\equiv m_3 r^3+O(r^4)$ with $m_3>0$ a constant,

\begin{eqnarray}\label{mupn3}
\mu=-p_z&=&\lambda_2= 6 m_3 \cos^4 (\psi/2),\\
p_x=p_y&=&-\lambda_1=3 m_3 \cos^2 (\psi/2) (\cos \psi-3).\nonumber
\end{eqnarray}

\begin{figure}[ht]
\includegraphics[scale=.7]{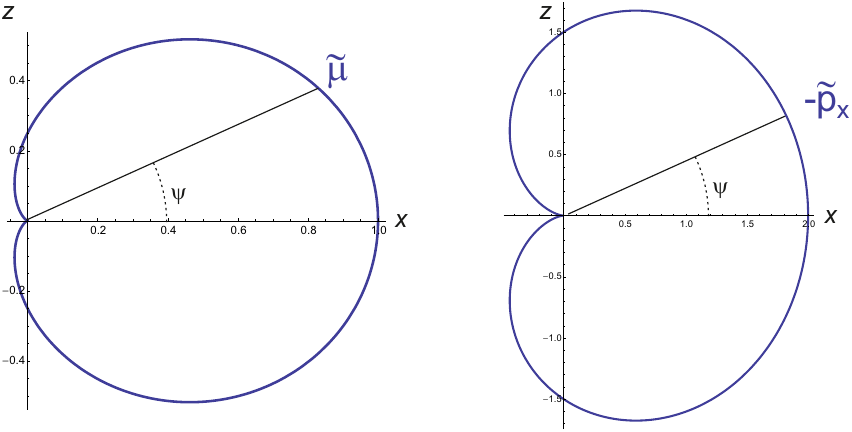}
\caption{\label{denspres} Polar plot of $\tilde\mu\equiv \mu/(6 m_3)$ and (minus) $\tilde p_x\equiv p_x/(3 m_3)$ around the ring for the case $\mathcal M(r)\sim r^3$ around $r=0$.}
\end{figure}

The effective density is non-negative while the effective pressures are all negative around the ring. These negative effective pressures imply that the weak energy conditions (WEC) are violated near the ring. Nevertheless, this is not surprising since all regular RBH violate the WEC, as shown in \cite{TorresReg}. For both the effective density and pressures, the value zero is reached in the direction of the disk $\psi=\pi/2$ and the maximum absolute value is reached in the opposite direction  $\psi=0$. Polar plots of the effective density and pressures for this particular case are shown in figure \ref{denspres}.

%v2
It is interesting to note that in the spherically symmetric case it has been argued that the core of the regular black hole should be De Sitter-like (see, for instance, \cite{Dymni} and references therein) admitting an effective energy-momentum tensor around $r=0$ of the type $T_{\mu\nu}\simeq -\Lambda g_{\mu\nu}$ and a mass function $\mathcal M(r)\simeq \frac{\Lambda}{6} r^3$. In the rotating case some authors call the core of the black hole \textit{De Sitter} when the mass function is of the order $O(r^3)$ around $r=0$. Nevertheless, it should be noted that the effective energy-momentum would only be strictly De Sitter-like if $\lambda_1=\lambda_2\neq 0$.  From the definitions of $\lambda_1$ (\ref{lambda1}) and $\lambda_2$ (\ref{lambda2}) and for their expressions around the ring in the case $n=3$ (\ref{mupn3}) it is clear that the disk (and the ring when approached with $\psi=\pi$) can be better described as \textit{minkowskian} ($\mu=p_x=p_y=p_z=0$). Only observers approaching the ring from outside and in the plane $x-y$ ($\psi=0$) would measure $\lambda_1=\lambda_2 (=6 m_3)$.
%
%Note that this condition represents a differential equation for $\mathcal M(r)$ in which $\theta$ appears. It follows %that is has no solution unless $a=0$, i.e. only in the spherically symmetric case (as stated before), but not for the %rotating case.
%v2end

\section{Causality}\label{secCaus}

In general, it seems reasonable to ask a time orientable spacetime to be absent of closed causal curves since the existence of such curves would seem to lead to logical paradoxes.
A spacetime devoid of closed causal curves is said to be \textit{causal} \cite{H&E}. If, in addition, no closed causal curve appears even under any small perturbation of the metric the spacetime is called \textit{stably causal}.
It is well-known that the Kerr metric with the usual extension allowing regions with negative $r$ is non-causal. Since Kerr metric is a particular case of the metric (\ref{gIKerr}), it is natural to ask whether regular RBH should also be non-causal.

Along the lines in \cite{Maeda}, in order to examine this issue we will use proposition 6.4.9 in \cite{H&E} which states that when a time function $f$ exists in the spacetime such that its normal $\mathbf{n}\equiv\nabla_\mu f \, dx^\mu$ is timelike, then the spacetime is stably causal. ($f$ can be thought as \emph{the} time in the sense that it increases along every future-directed causal curve).
Let us choose the time coordinate $\tilde{t}$ in Kerr-Schild coordinates as our time function $f$. The timelike character of $\mathbf{n}$ can be checked as follows:
\begin{equation}\label{n2}
\mathbf{n}^2=g^{\mu\nu} \nabla_\mu \tilde{t} \nabla_\nu \tilde{t}=g^{\tilde{t}\tilde{t}}=-1-\frac{2 \mathcal M(r) r^3}{r^4+a^2 z^2}.
\end{equation}
Since we would like this to be negative, it trivially follows
\begin{prop} \cite{Maeda}
If $r \mathcal M(r)\geq 0$ for all $r$, then the model of RBH with metric (\ref{GGg}) [or (\ref{gIKerr})] will be stably causal.
\end{prop}
Note that for a regular RBH (unextended through $r=0$), it just suffices to guarantee a non-negative mass function for the spacetime to be stably causal.

\section{Global structure}\label{Horizons}

Since regular RBHs do not need an extension with regions where $r$ takes negative values, the global structure of these spacetimes will necessarily differ from the global structure of the maximally extended Kerr spacetimes.

In order to get the global structure of regular RBHs we would need to locate their null horizons. In the general RBH case we should solve
\begin{equation}\label{eqdelta}
\Delta=r^2-2 \mathcal M(r) r+a^2=0.
\end{equation}
Without the knowledge of a specific $\mathcal M(r)$ it is not possible to know the \emph{exact} position of the horizons. Nevertheless, one can analyze the general behaviour of the horizons by taking into account the following considerations:
\begin{itemize}
\item If we assume an asymptotically flat spacetime, at large distances $\mathcal M(r)\simeq m=$constant, so that one (approximately) recovers the behaviour for the Kerr solution. Then $\Delta>0$ and $r$ will be a spacelike coordinate.
    %v2
    This is a usual assumption for black holes in the absence of a cosmological constant. For example, one can easily check that this is the case for the Bardeen and Hayward black holes (\ref{B&H}).
    %v2end
%
\item For $r\simeq 0$ ($a\neq 0$) a regular RBH has $\Delta>0$ thanks to the effect of the rotation and, again, $r$ will be a spacelike coordinate. (Note that this already happens in the classical Kerr solution).
\item If we assume the existence of a RBH and, thus, the existence of an exterior horizon $r_+$ (solution of $\Delta=0$) then the continuity of $\Delta$ and the two previous items imply either a single horizon (\textit{extreme} RBH), two horizons $r_-$ and $r_+\ (>r_-) $ or, in general, an even number of horizons.
\item If no solutions of (\ref{eqdelta}) exist, then no null horizons exist and we are in a \textit{hyperextreme} case. The regular rotating astrophysical object without an event horizon is not properly a black hole. The regularity implies that, contrary to the classical case, there is not a naked singularity.
\end{itemize}

In practice, the usual regular RBH in the literature has one or two null horizons, as in the classical case. This is not surprising if one considers deviations from General Relativity as coming from Quantum Gravity effects. Then, based on a simple dimensional analysis, one could expect the Planck scale to be the most natural scale in which to expect the departure from General Relativity to occur, which would imply only strong deviations from the classical solution around $r\sim r_{Planck}$ and, thus, only small corrections to the horizons (at least for RBHs with masses much larger than the planckian mass). One also expects that associated with non-singular RBHs there would be a \textit{weakening of gravity}. An effect which should be very important at high curvature scales. In this way, comparing with the classical case, it is usual to obtain bigger inner horizons and smaller outer horizons.
Of course, the Planck scale approach could turn out to be too naive and bigger deviations from the classical solutions could be possible, what would be good news for the observational aspects of RBHs.

\begin{figure}[htp]
\includegraphics[scale=.7]{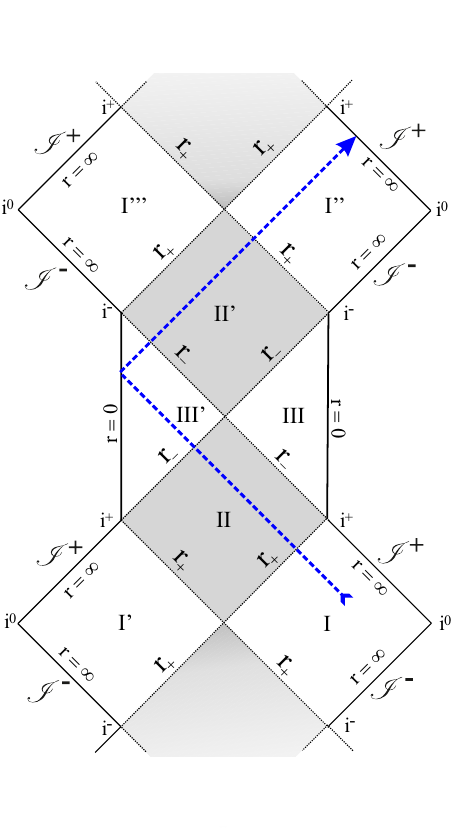}
\caption{\label{Pbiggera3} Penrose diagram for a regular rotating black hole with two null horizons. An extension through $r=0$ is not required and, thus, $r\geq 0$ in the maximally extended spacetime. The grey regions are the regions where the coordinate $r$ is timelike. We have depicted a light-like geodesic (dashed blue line) that, starting from the asymptotically flat region I, enters region II by traversing the \emph{event horizon} $r_+$. Then it reaches region III' by traversing the null horizon $r_-$. The value of $r$ first decreases along the geodesic until reaching $r=0$, where it increases again. It makes it to another null horizon $r_-$, enters region II', traverses another event horizon $r_+$ to enter the asymptotically flat region I'' where it travels towards the future null infinity. (Note that, since there are not singularities, the diagram is valid for all $\theta$). }
\end{figure}

\begin{figure}[htp]
\includegraphics[scale=.7]{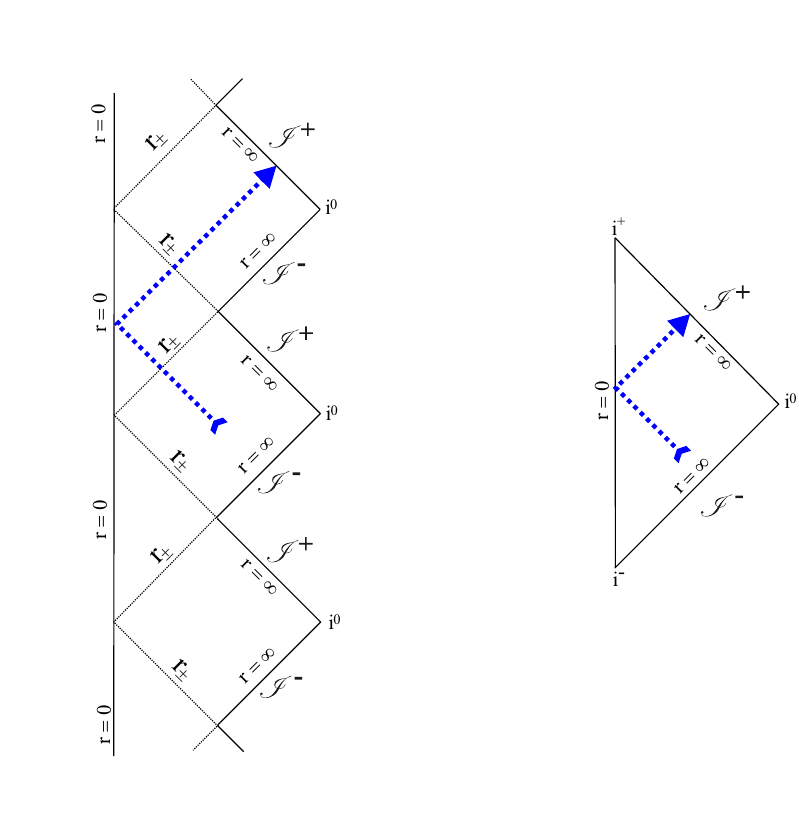}
\caption{\label{PHE3} Penrose diagrams for an extreme regular rotating black hole (left) and for a hyperextreme case (right). In both cases, an extension through $r=0$ is not required and $r\geq 0$. In the extreme case there is only one horizon denoted by $r_\pm$ in which the coordinate $r$ is lightlike. $r$ is never timelike. $r_\pm$ acts as an event horizon.
 In the hyperextreme case there are no horizons and $r$ is always spacelike. In both diagrams a light-like geodesic (dashed blue line) travels towards $r=0$, traverses the disk (or ring) and continues its travel towards the future null infinity. (Note that, again, since there are not singularities, the diagrams are valid for all $\theta$).}
\end{figure}

Nevertheless, in order to illustrate the global causal structure of regular RBHs let us follow the approach of small perturbations with respect to the classical horizons.
In this way, we can have three possible qualitatively different causal structures for the BH spacetime which are represented in the Penrose diagrams of figure \ref{Pbiggera3} (for the case with two null horizons) and of figure \ref{PHE3} (for the \textit{extreme} case and the \textit{hyperextreme} case).

The absence of an event horizon in the hyperextreme case is interesting, since this implies that an observer could receive information from the inner high curvature regions near $r=0$. In principle, this could be used to observationally test the different approaches to Quantum Gravity.
The problem is whether such RBHs are feasible. In the framework of General Relativity, it does not seem possible to obtain such high speed RBH ($a^2>m^2$) from a collapsing star and any attempt to overspin an existing black hole destroying its event horizon has failed, in agreement with the weak cosmic censorship conjecture \cite{Penrose}. However, for regular RBHs it has been suggested that it could be possible to destroy the event horizon \cite{Li&Bambi}.

\section{Conclusions}\label{conclu}

Under suitable conditions, the collapse of an astrophysically significant body can generate a black hole. Since one expects the generator of the black hole to be a rotating body, the black hole will rotate. General Relativity provides us with solutions for rotating black holes and predicts their characteristics, which are compatible with current observations. Nevertheless, the existence of inner singularities in the classical solutions for RBHs and the fact that General Relativity is incompatible with Quantum Mechanics leads us to search for better singularity-free models for RBHs, often based in some approach to a Quantum Gravity Theory.

Assuming that a manifold endowed with its corresponding metric is a fairly good approximation for describing regular RBHs, most of the models in the literature are of the Gürses-Gürsey type.
Remarkably, the analysis of these regular RBHs lead us to conclude that regular RBHs have a high degree of differentiability at the disk (at least $C^n$, with $n \geq 3$). In this way, in contrast to the classical Kerr RBH, regular RBH do not have differentiability problems at the disk, Christoffel symbols and the extrinsic curvature of the disk are well-behaved and an extension through $r=0$ is not needed. As a consequence, causality problems could be avoided simply if the mass function of the regular RBH is non-negative (which seems a desirable property of the mass function in most cases).

With regard to the ring, it has been shown that it is devoid of conical singularities. The effective density and pressures vanish at the ring if $\mathcal M (r)\sim r^n$ with $n>3$ around $r=0$. By contrast, in the particular case $\mathcal M (r)\sim r^3$ around $r=0$ one expects observers to measure a jump in the (finite) effective density and pressures (with a specific directional character) while crossing the ring. Note that this is loosely similar to the situation that an observer detects when crossing a star's surface modelled by matching, for instance, an exterior vacuum field with an interior fluid \cite{Seno}. 

The fact that the metric written in K-S coordinates is Lipschitz continuous ($C^{1-}$) around the ring for $n=3$ represents a nuisance when analyzing the geodesics crossing the ring in this case. Nevertheless, it can be shown that these geodesics exist and that they have continuous velocities ($C^1$ curves)\cite{Stein}\cite{Samann&S}.\footnote{Similar conclusions can be reached by applying the approach of \textit{optimal regularity}. (See \cite{RMTB}\cite{RMTB2} and references therein).}

%There appears a difficulty in the $n=3$ case if one tries to use K-S coordinates to describe the geodesics that do cross the ring (due to the lack of differentiability of %the metric at the ring in these coordinates). The good physical behaviour at the ring suggests that the proper study of geodesics crossing the ring in the $n=3$ case needs %a coordinate change presumably following the procedures in \cite{RMTB} and references therein.

%It implies that in the $n=3$ case there exists a local coordinate system in which the metric will be just $C^1$ at the %ring.

In summary, contrary to classical RBH solutions, regular RBH avoid singularities, (consequently) the violation of the cosmic censorship conjecture \cite{Penrose}, the existence of regions where the mass function acts repulsively and causality violations.

\section*{Acknowledgements}
I would like to thank Prof. Moritz Reintjes for our discussions on optimal regularity.

\end{document}